\documentclass[a4paper,11pt,oneside]{article}

\usepackage[english]{babel}
\usepackage[T1]{fontenc}
\usepackage[utf8]{inputenc}

\usepackage[pdftex,unicode]{hyperref}
\hypersetup{pdftitle=Economic Conditions for Innovation: Private vs. Public Sector}
\hypersetup{pdfauthor=Tomáš Evan and Vladimír Holý}

\usepackage[margin=60pt]{geometry}
\usepackage{amsmath}
\usepackage{amssymb}
\usepackage{bm}

\usepackage{amsthm}
\newtheorem{hypothesis}{Hypothesis}

\usepackage[group-minimum-digits=3]{siunitx}

\usepackage{enumitem}

\usepackage{graphicx}
\usepackage{float}
\usepackage{hhline}
\usepackage{multirow}
\usepackage{rotating}
\usepackage{booktabs}
\usepackage{dcolumn}

\usepackage[authoryear]{natbib}

\begin{document}

\begin{center}
{\Large \bfseries Economic Conditions for Innovation: \\ Private vs. Public Sector}
\end{center}

\begin{center}
{\bfseries Tomáš Evan} \\
Czech Technical University in Prague \\
Thákurova 2077/7, 160 00 Prague 6, Czechia \\
\href{mailto:tomas.evan@fit.cvut.cz}{tomas.evan@fit.cvut.cz}
\end{center}

\begin{center}
{\bfseries Vladimír Holý} \\
Prague University of Economics and Business \\
Winston Churchill Square 1938/4, 130 67 Prague 3, Czechia \\
\href{mailto:vladimir.holy@vse.cz}{vladimir.holy@vse.cz} \\
\emph{Corresponding Author}
\end{center}

\begin{center}
{\itshape \today}
\end{center}

\noindent
\textbf{Abstract:}
The Hicks induced innovation hypothesis states that a price increase of a production factor is a spur to invention. We propose an alternative hypothesis restating that a spur to invention require not only an increase of one factor but also a decrease of at least one other factor to offset the companies' cost. We illustrate the need for our alternative hypothesis in a historical example of the industrial revolution in the United Kingdom. Furthermore, we econometrically evaluate both hypotheses in a case study of research and development (R\&D) in 29 OECD countries from 2003 to 2017. Specifically, we investigate dependence of investments to R\&D on economic environment represented by average wages and oil prices using panel regression. We find that our alternative hypothesis is supported for R\&D funded and/or performed by business enterprises while the original Hicks hypothesis holds for R\&D funded by the government and R\&D performed by universities. Our results reflect that business sector is significantly influenced by market conditions, unlike the government and higher education sectors.
\\

\\

\\

\section{Introduction}
\label{sec:intro}

The goal of this paper is to analyse and verify the induced innovation hypothesis of J.R. Hicks first published in his Theory of Wages in 1932, which attributes a spur to invention to a price increase of a production factor \citep{Hicks1963}. This hypothesis was mainly tested on wages and their impact on labour saving technologies and, more recently, the impact of high energy prices on environmental technology innovations and energy savings.

In our preliminary study \citet{Bolotov2017}, we have attempted to falsify, in the sense of Popper, the Hicks hypothesis by means of orthodox modelling. Using regression model for intellectual property protection as the proxy variable for innovation, we have found that an increase in relative price of one factor mandates a relatively low price level of the other factor(s) to offset the companies' cost, for the innovations to take place. While the low cost of other factor(s) of production does not diminish the motivation to substitute the high-priced factor of production, it also gives companies the necessary capital for the innovation process. We follow our preliminary work and investigate this alternative hypothesis from two perspectives.

First, we review Hicks hypothesis and demonstrate the need for our alternative hypothesis in historical examples with focus on the industrial revolution in the United Kingdom. The clearest example of the phenomena at hand is that the industrialization needed two conditions to get started, that is, high wages and low prices of coal, as proven by economic historians such as \cite{Botham1987}, \cite{Allen2011} and \cite{Kelly2014}. As a relatively low price of the second factor of production had to be present, the Hicks original hypothesis is of no use for the second largest economic transformation after the introduction of agriculture.

Second, we theoretically reflect the changing conditions for the innovative process in mixed economies surrounding us in opposition to liberal economies before 1945 or even 1914 and empirically test relationships among the key variables influencing the innovation process in developed countries. As a measure of innovation activities, we use the investment into research and development (R\&D). It is universal for all industries and sectors and allows us to clearly determine the link between innovation and private and public sources of its financing. For detailed analysis of the relationship between the R\&D investment and the innovation performance, see e.g.\ \cite{Savrul2015} and \cite{Baumann2016}. As drivers of innovation, we consider wages and the price of oil (which replaced coal as a main source of energy). It is well documented in the literature that wages have positive effect on innovations (see e.g.\ \citealp{Bogliacino2018}). Concerning energy prices, \cite{Popp2002}, \cite{Crabb2010}, \cite{Verdolini2011} and \cite{Ley2016} find that their effect on energy-efficient innovations is also positive. In our study, however, we show that the effect of energy prices on innovations (not necessarily aimed at energy efficiency) may differ in relation to the public and private sector. Specifically, we analyze investments to R\&D over all industries broken down by the source of funding and the sector of performance in 29 OECD countries from 2003 to 2017. This detailed sector breakdown and generality for all industries are the main characteristics distinguishing our study from the others. Our main finding is that innovations funded and/or performed by the business enterprise sector are positively driven by wages and negatively by the oil price. Unlike in the other studies, high energy prices here acts as a lack of resources for innovations rather than the motivation for energy-efficient innovations. The government and higher education sectors are unaffected by energy prices and their innovations are driven just by high wages. 

The rest of the paper is structured as follows. In Section \ref{sec:orig}, we review the original Hicks hypothesis. In Section \ref{sec:alt}, we demonstrate the need for our alternative hypothesis. In Section \ref{sec:inn}, we discuss potential variables involved in contemporary innovation activities. In Section \ref{sec:data}, we introduce the analyzed data sample. In Section \ref{sec:model}, we build panel regression models relating R\&D expenditures to the economic environment. In Section \ref{sec:res}, we present results of the panel regression and assess the original Hicks hypothesis with our alternative. We conclude the paper in Section \ref{sec:con}.

\section{Hicks' Theory of Induced Innovation, Proofs and Criticism}
\label{sec:orig}

Numerous studies of the history of economics, including those in recent years, such as \citet{Lee2007}, \citet{Angelini2009}, \citet{Gallouj2009}, \citet{Savona2013}, \citet{Fabre2014}, and \citet{Milyaeva2015}, have consistently shown that innovations benefit companies, industries and economies in terms of increasing competitiveness, economic growth and development. There is however little consensus on what the main causes of innovation are. Sir J.R. Hicks \citep[p.~124]{Hicks1963} has stated the following hypothesis, which later became the foundation of Hicks' widely discussed Theory of Induced Innovation.

\begin{hypothesis}[Hicks Induced Innovation Hypthosis]
\label{hyp:hicks}
A change in the relative prices of the factors of production is itself a spur to invention, and to invention of a particular kind -- directed to economizing the use of a factor which has become relatively expensive.
\end{hypothesis}

The relative straightforwardness and immense implications of the Induced Innovation Theory together with the name of the well-known British economist and Nobel Prize laureate has caused the theory to be widely discussed from the moment it was formulated. The theory had both proponents (\citealp{Fellner1961, Fellner1971}, \citealp{Samuelson1965}, \citealp{Kennedy1967}, i.a.) as well as opponents. The latter include Nobelist W.D. Nordhaus who criticised it for necessitating very strong and limiting preconditions, among other flaws. Nordhaus thinks “the model is too defective to be used in serious economic analysis” \citep[p.~208]{Nordhaus1973}. Others criticised the lack of economic foundations which are implicit in the theory and tried to establish them \citep{Funk2002}.

The crucial fact that the production factor shares stay relatively constant in the production function remains a widely accepted stylized fact. Yet, whether this is caused by instant “spur to invention” and thus serves as proof of the theory, remains as controversial as it was back in Hicks' time. Perhaps even more controversially, there is still generally accepted mechanism by which changes in factor prices affect inventive or innovative activity (\citealp[pp.~43--44]{Salter1966}, \citealp{Ahmad1966}, \citealp{Hayami1971}, \citealp{Ruttan1984}, i.a.).

Not even the theory's critics can deny, however, that it has been used and in quite a few cases proven empirically. Attempted from the start and followed by a groundbreaking paper of William Fellner “Empirical Support for the Theory of Induced Innovation” \citep{Fellner1971}, there have been several fields in which endorsement and application could be found. The line of research empirically confirming the main hypothesis included in the theory was originally centred on high wages spurring labour-saving innovation, and later agricultural development. More recently the emphasis has shifted towards energy prices and induced innovation in energy-saving technologies.

\citet{Newell1999}, i.a., found that the rate of overall innovation is independent of energy prices and regulation. The direction of innovation, however, was responsive to energy price changes for several products tested by the authors. \citet{Popp2002}, using patent citations as a measure of supply of knowledge, found that both energy prices and the quality of existing knowledge have significantly strong positive effects on innovation. Also using patent counts and citation data, \citet{Jang2013} confirm that demand and supply factors -- including knowledge stocks and crude-oil price -- have positive and statistically significant effects on technological biofuel innovations in the United States of America.

There is, however, also a relatively large number of other correlates to innovation such as inward foreign direct investment, outward foreign direct investment, imports, state guarantees and incentives among many other, as stressed by \citet{Lin2008}. Explaining the causes of innovation is therefore a long-standing problem in social science, while the large body of existing literature has not been conclusive. This paper adds to this ongoing discussion an attempt to alter significantly the existing Hicks hypothesis to the point of its negation, among others ways through evidence from economic history, specifically, from the example of industrialization in the U.K. (the initial one) and in the world.

\begin{table}
\centering 
\caption{Take-off dates of \citet{Rostow1991}.}
\label{tab:takeoff}
\begin{tabular}{@{\extracolsep{5pt}}ll} 
\\[-1.8ex]\hline 
\hline \\[-1.8ex] 
Country & Take-Off \\ 
\hline \\[-1.8ex] 
United Kingdom & 1783--1802 \\
France         & 1830--1860 \\
Belgium        & 1833--1860 \\
United States  & 1843--1860 \\
Germany        & 1850--1873 \\
Sweden         & 1868--1890 \\
Japan          & 1878--1900 \\
Russia         & 1890--1914 \\
Canada         & 1896--1914 \\
Argentina      & 1935--     \\ 
Turkey         & 1937--     \\
India          & 1952--     \\
China          & 1952--     \\
\hline 
\hline \\[-1.8ex] 
\end{tabular} 
\end{table}

\section{Historical Examples of Hicks' Theory and the Need for an Alternative Hypothesis}
\label{sec:alt}

One indication that the increase in relative price of one factor mandates a relatively low price level of other factor(s) of production would be to look at historical examples of eras of rapid innovation \footnote{This will allow us to determine cause and effect, whether innovations are spurred by changes in relative prices of inputs with the motive to economize or not.}. A near-perfect example of such rapid and continuous innovation is the industrial revolution. There were many causes and necessary conditions for this long and complicated process of introduction of mechanized production. For Rostow's daring attempt to assign dates to various countries' “take-offs”, see Table \ref{tab:takeoff} (\citealp[Chapters 3--4]{Rostow1991} and \citealp{Baldwin1999}). These included political stability, sufficient capital accumulation, relatively mature banking sector, etc. (for a somewhat more comprehensive list, see the works of \citealp[Chapters 13--16]{Landes1998}, \citealp[Chapters 5--6]{Pomeranz2000}, \citealp[Chapters 2, 3, 6]{Maddison2007}, \citealp[Chapters 1--2]{Evan2014}, as well as others).

Apart from these uncontroversial conditions which were sooner or later fulfilled in most countries of the northern hemisphere at least, there is still the issue of the cause of this far-reaching economic and social change. Particularly the question, to quote \citet{Allen2009, Allen2011, Allen2015}, “why the industrial revolution was British?” In other words, why was the innovation in Great Britain spurred fifty or more years earlier (end of the 18th century) than in nearby countries with similar socio-economic characteristics? There is hardly any discussion about the prime reason for businessmen trying to replace human labour with machines. The reason is high wages, for calculations see \citet{Botham1987} and \citet{Kelly2014}. While not uniquely high, since both GDP and income per capita were higher still in the Netherlands, British wages were the prime incentive for the innovation of labour-saving techniques. This, however, would not be enough, as it was not enough in the Netherlands which became industrialized much later than Great Britain, as shown by \citet[Chapters 15–16]{Landes1998}. A relatively low price of the second factor of production had to be present. The second factor in the case of British industrialization was cheap energy in both the textile and iron industries. Such energy was ensured by fast running streams for textile mills at first, and soon replaced by accessible, abundant coal of good enough quality. Thus, the high-wage economy of London together with the use of cheap coal shipped in from Newcastle led to the early industrialization of Britain, as it motivated mechanical production and allowed this innovation by savings made from cheap energy.

Had there not been both factors of production in this fortunate price combination, the first industrialization would not have been British, but Dutch or German perhaps, see \citet[p.~366]{Allen2011}, also \citet[Chapters 15–16]{Landes1998} and \cite{Allen2015}\footnote{To sum up, a high price of one factor offset by a low price of another input could be sufficient to spur innovation in the U.K., which is opposite logic to the effect described by Hicks.}. 

Energy prices were dominant for the iron and steel industry in Britain. For textiles the costs of inputs of raw materials were even more important. For many centuries Britain exported wool and woollens and to a lesser extend linen. While making a good export product and enriching both land owners and merchants there was no way how to accumulate enough capital to mechanize production, nor was there a reason to do so. It was only after cotton became available, the price of which could be forced down to levels constituting a significant advantage, that the combination of high wages of spinners and weavers together with cheap cotton from the West Indies and later plantations in the American South allowed for induced innovation, as shown by \citet{Broadberry2013}, and \citet{Tomory2016}. This combination was so powerful that British industrialists overcame a number of obstacles including strong competition from high quality Indian cotton products as well as the fact that all raw cotton had to be imported from faraway and often unstable locations. 

Once started, the innovative process reinforced itself in several ways. It allowed for tremendous economies of scale and profits trumping all other non-mechanized productions around the world. This, in turn, put pressure on wages to keep rising, thus motivating the implementation of further labour-saving techniques. The industrial revolution created so called Big Divergence (see \citealp{Pritchett1997}) increasing incomes in industrialized countries (Europe, USA, Japan) and creating “innovative centres” in these countries while de-industrializing everyone else (see \citealp{Landes1998}, \citealp{Maddison2007}, etc.).

The industrial revolution provides us with arguably the clearest historical example of the phenomena, as the impact of innovations during other periods such a World War II is more influenced by accompanying political or military events. Oil crises of 1973 and 1979 also provide seemingly contra-intuitive and definitely contra-Hicksian example.  Despite increasing relative wages AND increasing relative prices of energy which should lead to drastic spur to innovation, the inventive input actually decreased significantly from 60s to 70s puzzling researchers even a decade later \citep{Griliches1989}.

The described historical events above clearly demonstrate the need to extend the Hicks' hypothesis. To do so, we formulate the following alternative hypothesis to the Hicks' theory.

\begin{hypothesis}[Alternative Induced Innovation Hypothesis]
\label{hyp:alt}
Innovation is spurred by an increase in the relative price of one factor of production compensated by a decrease in relative price of another factor of production.
\end{hypothesis}

\section{Wages, Oil Prices and Government Policies Fuelling Innovation}
\label{sec:inn}

This paper attempts to reflect the changed conditions for innovative process in societies surrounding us today. The current economic conditions are not of market economy but mixed economy instead, while government finances and guarantees much of the research and development (R\&D) in developed countries around the world (see Table \ref{tab:gerdShare} and Figure \ref{fig:gerdCountryFund}). It is unlikely, therefore, that the conditions motivating for the innovation would remain the same as were in previous two centuries. In general, the government is one of the determinants for innovation capacity, but the size and effectiveness of its involvement is highly debatable.  \cite{Wang2018a} concludes on examples from Singapore and Hong Kong that innovation can be increased by both strong government intervention focused on big players in the former and with minimal government activity providing an environment for small firms innovation as in the latter. Regarding the ideology of government most fitting for innovation \cite{Wang2019} claim leftist ruling party limits technical innovation, whereas a right-wing ruling party promotes the appearance of new technology. Some authors suggest government innovation is particularly needed for R\&D intensive sectors. \cite{Yigitcanlar2018}, inter alia, claim critical importance of public innovation funding.  In the case of Brazilian software companies, they prove that companies using public funding are more likely to become nationally and internationally competitive as opposed to the companies using commercial banks as a source of their financing.

Therefore, in this paper, we try to identify whether the market conditions of relative prices of factors of production still hold sway as motivating factor for the spur to innovation, according to our alternative Hicks hypothesis. Or, if the motivation is more in line with the motivation that can be expected from the governmental sector. This would include correlation of research output with budgetary constraints and GDP levels, which is the base for government expenditure rather than any market conditions. We also include educational sector, that is, universities, as well as non-governmental research oriented institutions in our analysis. The last included actor with potentially significant impact on research and development levels of a country would be foreign based businesses out of which those using foreign direct investment as a mode of entry for their investment might be of relevance for country's R\&D. The extend of this influence needs more study as multinational corporations tend to rely more on internal sources of financing in comparison with their other operations as the poor institutions in the host country make the external financing of R\&D costly \citep{Alam2019}.

In line with our previous research and concerning literature we consider impact of relative prices of different factors of production and keep wages and price of oil as the two most relevant. Oil has clearly replaced coal as the most important source of energy while labour costs remain the single most important factor of production across industries. There is large body of literature relating wages of skilled and unskilled labour and innovation. It is clearly two-way street with high wages motivating innovation and innovation favouring high-skilled labour (\citealp{Bogliacino2018}, i.a.). In the light of our results somehow surprisingly the available literature considers energy prices as to have strong and positive effect on innovation. \cite{Popp2002} suggests the impact is so strong it can be advantageous for government to use market-based environmental policies to help ameliorate global warming with the help of price-induced technological change. In several sectors of economy (for a review see \citealp{Ruttan2001}) higher oil prices were found to lead to an increased innovation. Particularly in the automotive industry as it is an energy-intensive sector. \cite{Crabb2010} also suggest, however, a government intervention, this time based upon the premise of possible under-investment in private research and development given the relatively slow diffusion of knowledge in the sector combined with high effectiveness of carbon-based taxes in encouraging innovation among other things. Somehow opposite view can be found on European electricity industry where more deregulation namely in barriers of entry, public ownership and vertical integration seems to have positive impact on innovation activity \citep{Cambini2016}.

The innovation process can be divided into the R\&D stage and the commercialization stage (see e.g.\ \citealp{Zhang2019, Wang2020}). As a measure of innovation activities, we utilize the size of investments to R\&D as it can be directly attributed to the private or public sector. The investment is of course just the beginning of the innovation process but is a clear and universal indicator of its magnitude. On the other hand, measurable R\&D outputs such as the number of patents, the share of high-technology exports and the number of scientific publications relate only to specific industries and sectors. \cite{Savrul2015} and \cite{Baumann2016} study the link between the R\&D investment and the innovation performance. \cite{Holy2018e} investigate the relation between the inputs and outputs of the R\&D process and find that countries with higher GDP per capita may not necessarily produce more outputs in terms of the patent and citation counts, but are significantly more efficient in transforming the investments and the human capital into these outputs.

\section{Analyzed Sample of OECD Countries}
\label{sec:data}

\begin{figure}
\begin{center}
\includegraphics[width=0.9\textwidth]{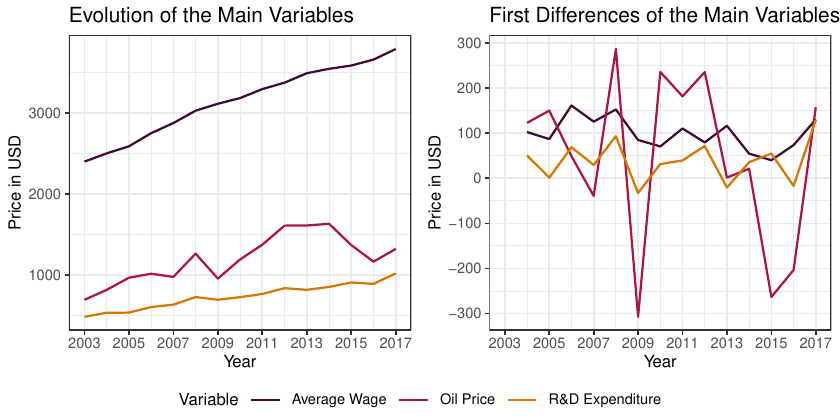}
\caption{The total R\&D expenditure per capita, average wage and oil price averaged over 29 OECD countries.}
\label{fig:varTime}
\end{center}
\end{figure}

We investigate the member countries of the Organisation for Economic Co-Operation and Development (OECD) from 2003 to 2017. The main analyzed variable is the annual gross domestic R\&D expenditure per capita. The economic environment is represented by the average monthly wage and automotive diesel oil price per \num{1000} litres. All variables are current prices in USD adjusted for purchasing power parities. The evolution of the variables over time is illustrated in Figure \ref{fig:varTime}. The source of the R\&D expenditures and average wages is OECD while the source of the oil prices is the International Energy Agency (IEA). Additionally, as control variables, we utilize the annual gross domestic product (GDP) per capita, the inflation rate, international trade in goods and services as percentages of GDP and foreign direct investment (FDI) flows as percentages of GDP from OECD as well as several Worldwide Governance Indicators from the World Bank (WB).

\begin{figure}
\begin{center}
\includegraphics[width=0.9\textwidth]{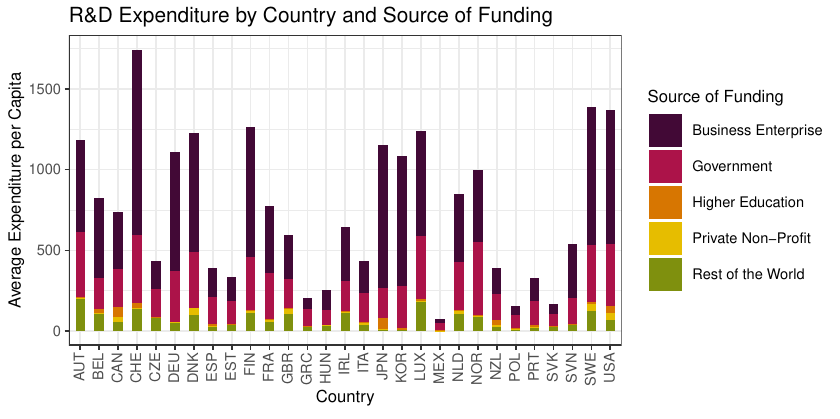}
\caption{The R\&D expenditures per capita averaged over time from 2003 to 2017 and broken down by the source of funding.}
\label{fig:gerdCountryFund}
\end{center}
\end{figure}

\begin{figure}
\begin{center}
\includegraphics[width=0.9\textwidth]{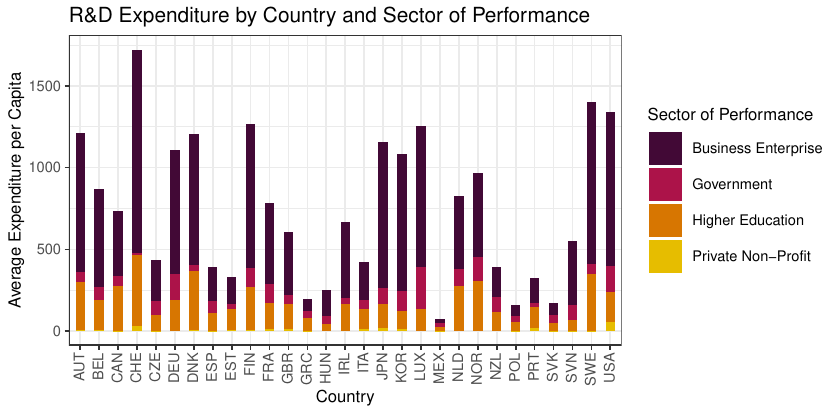}
\caption{The R\&D expenditures per capita averaged over time from 2003 to 2017 and broken down by the sector of performance.}
\label{fig:gerdCountryPerf}
\end{center}
\end{figure}

As the R\&D expenditure is our main object of interest, we further elaborate on it. The data are collected using the standard OECD methodology for statistics related to R\&D described in the Frascati Manual \citep{OECD2015}. Besides the total intramural gross domestic R\&D expenditures (Total), we also utilize the R\&D expenditures broken down by the source of funding and the sector of performance as well. There are five sources of funding: the business enterprise (Fund-BES), the government (Fund-GOV), the higher education (Fund-HES), the private non-profit (Fund-PNP) and the rest of the world (Fund-ROW). Furthermore, there are four sectors of performance: the business enterprise (Perf-BES), the government (Perf-GOV), the higher education (Perf-HES) and the private non-profit (Perf-PNP). The average shares of the R\&D expenditures for specific sources of funding and sectors of performance are shown in Table \ref{tab:gerdShare}. The dominant R\&D segment is the self-funded business sector with 52 percent share. Other significant segments are the higher education sector funded by the government with 17 percent share and the self-funded government sector with 11 percent share. Figure \ref{fig:gerdCountryFund} shows the composition of the source of funding in each country while Figure \ref{fig:gerdCountryPerf} shows the composition of the sector of performance.

Unfortunately, not all variables are available for all OECD countries. For this reason, we analyze only 29 of the 36 OECD member countries over 15 years. We exclude Australia, Chile, Iceland, Israel, Latvia and Lithuania from the analysis due to missing oil prices and Turkey due to missing average wages. Furthermore, our dataset contains some additional missing values. For the analysis of the total intramural R\&D expenditure, we have 382 observations with all variables. This means that 53 observations are missing. For the R\&D expenditures with a specific source of funding or sector of performance, we have between 334 and 385 observations. The exception is Fund-HES variable with only 293 observations and the Perf-PNP variable with only 285 observations. In these two cases, some countries are entirely missing due to differences in data collection methodology.

\begin{table}
\centering 
\caption{The shares of R\&D expenditures per capita in percents averaged over 29 OECD countries from 2003 to 2017.}
\label{tab:gerdShare}
\begin{tabular}{@{\extracolsep{5pt}}lrrrrrr} 
\\[-1.8ex]\hline 
\hline \\[-1.8ex] 
 & \multicolumn{6}{c}{Source of Funding} \\  
\cline{2-7} \\
[-1.8ex] Sector of Perf. & BES & GOV & HES & PNP & ROW & Total \\ 
\hline \\[-1.8ex] 
BES   &  51.80 &   4.50 &   0.04 &   0.13 &   6.51 &  62.98 \\ 
GOV   &   0.84 &  10.57 &   0.05 &   0.14 &   0.78 &  12.38 \\ 
HES   &   1.32 &  17.20 &   2.23 &   0.83 &   1.38 &  22.95 \\ 
PNP   &   0.25 &   0.70 &   0.01 &   0.53 &   0.19 &   1.69 \\ 
Total &  54.20 &  32.97 &   2.34 &   1.63 &   8.86 & 100.00  \\ 
\hline 
\hline \\[-1.8ex] 
\end{tabular} 
\end{table}

\section{Specification of Panel Regression Model}
\label{sec:model}

To analyze the influence of the economic environment on the R\&D expenditures, we utilize the panel regression. As the dependent variable, we consider the total intramural R\&D expenditure per capita as well as R\&D expenditures per capita with a specific source of funding and R\&D expenditures per capita with a specific sector of performance. Therefore, we build models with ten dependent variables in total. In all cases, we consider the average wage and the oil price as the independent variables.

In the preliminary step, we investigate dependence of the R\&D expenditures on various lagged values. Using auxiliary panel regression with independent variables lagged by one year, we find that the oil price Granger-cause the total R\&D expenditures and R\&D expenditures funded or performed by the business enterprise sector. Furthermore, in subsequent modeling, much stronger results are obtained when the oil price is lagged suggesting that there is a one-year delay before the decision to invest and the actual investment. For these reasons, we include the oil price lagged by one year in the final model. Concerning the average wage, the situation is not as straightforward. Slightly stronger results are obtained when using the current value instead of the lagged value in subsequent modeling, although the results with the lagged average wage are statistically significant as well. Nevertheless, we decide to utilize only the lagged values as the one-year lag in all independent variables allows for convenient temporal interpretation.

The considered specifications of the panel regression model are as follows. First, we find out whether individual and time effects are needed. For this purpose, we utilize the Lagrange multiplier test of \citet{Honda1985}. The p-values of the individual effects test for all dependent variables are virtually zero. In contrast, the p-value of the time effects test for the total intramural expenditure variable is 0.98 and is similarly high for all other dependent variables. Therefore, we include only individual effects in the model.

Next, we choose between the first-differences estimator and the within estimator. Both estimators deal with the fixed effects in the panel regression. Our variables are clearly non-stationary (see Figure \ref{fig:varTime}) and therefore we resort to the first-differences estimator as it removes any unit roots in the dependent and independent variables as well. In contrast, a spurious relation between the variables is a serious issue for the within transformation. The auxiliary-regression-based Hausman test of \citet[Section 10.7.3]{Wooldridge2010} suggests that random effects can be considered as well. However, the model with random effects would face the same issues as in the case of the within estimator.

Finally, we investigate the structure of the error terms. We adopt the heteroskedasticity test of \citet{Breusch1979} and the serial correlation test of \citet[Section 10.6.3]{Wooldridge2010} based on first differences. The test statistics are presented in Table \ref{tab:resFd} for the first-differences estimator and in Table \ref{tab:resFix} for the within estimator. We find that there is no universal behavior across our models in terms of heteroskedasticity and serial correlation. In general, however, the first-differences transformation removes serial correlation much better than the within transformation. This is another major motivation for the first-differences estimator. Furthermore, heteroskedasticity is present in many models for both estimators. To account for heteroskedasticity and remaining serial correlation in the error terms, we utilize the White method of \citet{Arellano1987} for robust estimation of the parameter covariance matrix.

The resulting panel model is as follows. Let $N$ denote the number of countries, $T$ the number of time periods and $M$ the number of independent variables. Further, let $y_{i,t}$ denote the dependent variable of country $i$ in time $t$ and $x_{j,i,t}$ the independent variable $j$ of country $i$ in time $t$. The linear panel model with first differences and lagged independent variables is then given by
\begin{equation*}
y_{i,t} - y_{i,{t-1}} = \beta_0 + \sum_{j=1}^M \beta_j \left( x_{j,i,t-1} - x_{j,i,t-2} \right) + e_{i,t}, \qquad i = 1,\ldots,N, \qquad t = 3,\ldots,T,
\end{equation*}
where $\beta_j$ are the unknown coefficients and $e_{i,t}$ are the error terms. Note that the first two time periods are used for inicialization of the lagged variables and first differences.

\section{Results and Implications}
\label{sec:res}

The estimated coefficients with standard deviations and summary statistics for the first-differences estimator are reported in Table \ref{tab:resFd}. Note that the R-squared statistic is relatively low for the first-differences estimator as it explains the change in the dependent variable. On the other hand, the R-squared is much higher for the within estimator as it explains the absolute value of the dependent variable which has a clear trend in time. Nevertheless, modeling changes is more meaningful in our situation while it also proves to be more challenging.

First, let us focus on the total intramural R\&D expenditures per capita (the Total model). The average wage has significantly positive effect on the R\&D expenditure in this model. Specifically, the annual R\&D expenditure per capita increases by 0.27 USD when the average monthly wage increases by 1 USD. The effect of the oil price is insignificant. The model explains 20 percent of the variance in the total R\&D expenditure per capita changes.

Next, let us consider breakdown by the source of funding. The dominant source of funding is the business enterprise sector with 54 percent share of the total R\&D expenditures on average. In the Fund-BES model, the average wage has significantly positive effect on the R\&D expenditure while the oil price has significantly negative effect. The annual R\&D expenditure per capita by the business enterprise sector increases by 0.22 USD when the average monthly wage increases by 1 USD. The annual R\&D expenditure per capita decreases by 0.03 USD when the oil price per \num{1000} litres increases by 1 USD. The Fund-BES model explains 16 percent of the variance. Another major source of funding is the government sector with 33 percent share of the total R\&D expenditures on average. In the Fund-GOV model, the average wage has significantly positive effect on the R\&D expenditure while the oil price has no significant effect. The annual R\&D expenditure per capita by the government sector increases by 0.10 USD when the average monthly wage increases by 1 USD. The Fund-GOV model explains 23 percent of the variance. Other sources of funding have lower share of the R\&D expenditure and are not explained well by our model due to very low R-squared statistic and insignificance of regressors.

Finally, let us consider breakdown by the sector of performance. The dominant sector of performance is the business enterprise sector with 63 percent share of the total R\&D expenditures on average. The behavior of the Perf-BES model is similar to the Fund-BES model although less pronounced. The annual R\&D expenditure per capita in the business enterprise sector increases by 0.17 USD when the average monthly wage increases by 1 USD. The annual R\&D expenditure per capita decreases by 0.01 USD when the oil price per \num{1000} litres increases by 1 USD. The Perf-BES model explains 11 percent of the variance. The second most important sector is the higher education sector with 23 percent share of the total R\&D expenditures on average. In the Perf-HES model, the average wage has significantly positive effect on the R\&D expenditure while the oil price has no significant effect. The annual R\&D expenditure per capita in the higher education sector increases by 0.06 USD when the average monthly wage increases by 1 USD. The Perf-HES model explains 13 percent of the variance. Other sectors of performance have lower share of the R\&D expenditure and are not explained well by our model due to very low R-squared statistic and insignificance of regressors.

As a robustness check, we also report results obtained by the within estimator in Table \ref{tab:resFix}. Just as with the first-differences transformation, we adopt the robust covariance matrix estimation of \citet{Arellano1987} to account for heteroskedasticity and serial correlation in the error terms. We find that there are no dramatical differences in the coefficients estimated by both methods. The within estimator, however, puts more significance to the average wage variable. Most likely, this is a spurious relation caused by non-stationarity. Overall, the within estimator does not notably deviate from the first-differences estimator but is not as reliable due to potential spurious relationship.

To further investigate robustness of our approach, we add several control variables to the model. We utilize nine additional variables capturing possible internal and external conditions for innovation -- the annual gross domestic product (GDP) per capita, the inflation rate, the Control of Corruption index, the Political Stability and Absence of Violence index, the Rule of Law index, the exports and imports of goods and services as percentages of GDP and the outward and inward foreign direct investment (FDI) flows as percentages of GDP. To be consistent with the previous models, we use the lagged values of all independent variables. The use of the current values of the control variables, however, leads to similar results in most cases. We estimate the extended model using the first-differences transformation and report results in Table \ref{tab:resAdd}. Note that missing values in some control variables further reduces the number of observations in the individual models. Overall, there are no major differences in the estimated coefficients of the average wage and oil price variables from the previous two models. Concerning added variables, only international trade and FDI flows in both directions proves to be relevant for some dependent variables.

To conclude, let us relate the results of the panel regression analysis to hypotheses \ref{hyp:hicks} and \ref{hyp:alt}. The original Hicks hypothesis is supported in the cases of R\&D funded by the government sector and R\&D performed by the higher education sector. In contrast, our alternative hypothesis is supported in the cases of R\&D funded by the business enterprise sector and R\&D performed by the business enterprise sector. The other R\&D expenditures are funded and performed on much lower scales and are not well captured solely by the average wages and the oil prices.

\begin{sidewaystable} \centering 
  \caption{The results of the panel regression based on the first-difference estimator.} 
  \label{tab:resFd} 
\footnotesize 
\begin{tabular}{@{\extracolsep{5pt}}lcccccccccc} 
\\[-1.8ex]\hline 
\hline \\[-1.8ex] 
 & \multicolumn{10}{c}{R\&D Expenditures} \\ 
\cline{2-11} 
\\[-1.8ex] & Total & Fund-BES & Fund-GOV & Fund-HES & Fund-PNP & Fund-ROW & Perf-BES & Perf-GOV & Perf-HES & Perf-PNP \\ 
\hline \\[-1.8ex] 
 Intercept & 7.4459 & $-$2.0300 & $-$0.4973 & 0.7228$^{**}$ & 0.5170$^{**}$ & 5.3768 & 5.6545 & 1.9911$^{**}$ & 2.4450 & $-$0.0445 \\ 
  & (11.1675) & (6.8087) & (2.0693) & (0.3186) & (0.2448) & (4.8662) & (8.1957) & (0.8659) & (2.1569) & (0.1983) \\ 
  & & & & & & & & & & \\ 
 Average Wage & 0.2671$^{**}$ & 0.2156$^{***}$ & 0.0998$^{***}$ & 0.0032$^{*}$ & 0.0033 & $-$0.0068 & 0.1690$^{*}$ & 0.0088 & 0.0640$^{***}$ & 0.0044$^{**}$ \\ 
  & (0.1214) & (0.0542) & (0.0174) & (0.0018) & (0.0030) & (0.0450) & (0.0894) & (0.0080) & (0.0238) & (0.0020) \\ 
  & & & & & & & & & & \\ 
 Oil Price & $-$0.0077 & $-$0.0328$^{***}$ & 0.0020 & $-$0.0010 & $-$0.0014$^{*}$ & 0.0118$^{*}$ & $-$0.0129$^{***}$ & 0.0034 & 0.0026 & $-$0.0019 \\ 
  & (0.0090) & (0.0088) & (0.0050) & (0.0015) & (0.0008) & (0.0068) & (0.0048) & (0.0028) & (0.0045) & (0.0012) \\ 
  & & & & & & & & & & \\ 
\hline \\[-1.8ex] 
Observations & 326 & 286 & 285 & 245 & 286 & 280 & 327 & 330 & 329 & 244 \\ 
R-Squared & 0.1997 & 0.1641 & 0.2330 & 0.0112 & 0.0270 & 0.0070 & 0.1138 & 0.0112 & 0.1262 & 0.0456 \\ 
Serial Correlation Test & 1.9062 & 15.8501$^{***}$ & 0.1198 & 1.0229 & 3.8027$^{*}$ & 0.5462 & 4.5562$^{**}$ & 0.04581 & 1.0832 & 1.7553 \\ 
Heteroscedasticity Test & 46.1395$^{***}$ & 7.3371$^{**}$ & 2.2193 & 24.5454$^{***}$ & 8.6088$^{**}$ & 18.1795$^{***}$ & 22.5090$^{***}$ & 6.3528$^{**}$ & 4.8959$^{*}$ & 0.5292 \\ 
\hline 
\hline \\[-1.8ex] 
\multicolumn{11}{r}{$^{*}$p$<$0.1; $^{**}$p$<$0.05; $^{***}$p$<$0.01} \\ 
\end{tabular} 
\end{sidewaystable}  

\begin{sidewaystable} \centering 
  \caption{The results of the panel regression based on the within estimator.} 
  \label{tab:resFix} 
\footnotesize 
\begin{tabular}{@{\extracolsep{5pt}}lcccccccccc} 
\\[-1.8ex]\hline 
\hline \\[-1.8ex] 
 & \multicolumn{10}{c}{R\&D Expenditures} \\ 
\cline{2-11} 
\\[-1.8ex] & Total & Fund-BES & Fund-GOV & Fund-HES & Fund-PNP & Fund-ROW & Perf-BES & Perf-GOV & Perf-HES & Perf-PNP \\ 
\hline \\[-1.8ex] 
 Average Wage & 0.3098$^{***}$ & 0.1743$^{***}$ & 0.0960$^{***}$ & 0.0082$^{***}$ & 0.0081$^{***}$ & 0.0472$^{***}$ & 0.1812$^{***}$ & 0.0284$^{**}$ & 0.1007$^{***}$ & 0.0041$^{***}$ \\ 
  & (0.0446) & (0.0397) & (0.0176) & (0.0023) & (0.0026) & (0.0067) & (0.0491) & (0.0132) & (0.0143) & (0.0016) \\ 
  & & & & & & & & & & \\ 
 Oil Price & $-$0.0181 & $-$0.0390$^{**}$ & $-$0.0059 & $-$0.0009 & $-$0.0030$^{*}$ & 0.0132 & $-$0.0096 & 0.0026 & $-$0.0113 & $-$0.0031 \\ 
  & (0.0268) & (0.0189) & (0.0096) & (0.0017) & (0.0016) & (0.0105) & (0.0253) & (0.0068) & (0.0112) & (0.0020) \\ 
  & & & & & & & & & & \\ 
\hline \\[-1.8ex] 
Observations & 355 & 315 & 314 & 271 & 315 & 309 & 356 & 359 & 358 & 266 \\ 
R-Squared & 0.7129 & 0.3838 & 0.5779 & 0.3290 & 0.3503 & 0.3435 & 0.4760 & 0.2885 & 0.7391 & 0.1085 \\ 
Serial Correlation Test & 1.9062 & 15.8501$^{***}$ & 0.1198 & 1.0229 & 3.8027$^{*}$ & 0.5462 & 4.5562$^{**}$ & 0.04581 & 1.0832 & 1.7553 \\ 
Heteroscedasticity Test & 4.1973$^{**}$ & 0.1171 & 1.0666 & 5.0044$^{**}$ & 5.3958$^{**}$ & 3.0722$^{*}$ & 1.2233 & 2.8058$^{*}$ & 1.0489 & 1.4447 \\ 
\hline 
\hline \\[-1.8ex] 
\multicolumn{11}{r}{$^{*}$p$<$0.1; $^{**}$p$<$0.05; $^{***}$p$<$0.01} \\ 
\end{tabular} 
\end{sidewaystable} 

\begin{sidewaystable} \centering 
  \caption{The results of the panel regression based on the first-difference estimator with additional control variables.} 
  \label{tab:resAdd} 
\footnotesize 
\begin{tabular}{@{\extracolsep{5pt}}lcccccccccc} 
\\[-1.8ex]\hline 
\hline \\[-1.8ex] 
 & \multicolumn{10}{c}{R\&D Expenditures} \\ 
\cline{2-11} 
\\[-1.8ex] & Total & Fund-BES & Fund-GOV & Fund-HES & Fund-PNP & Fund-ROW & Perf-BES & Perf-GOV & Perf-HES & Perf-PNP \\ 
\hline \\[-1.8ex] 
 Intercept & 8.7551 & $-$2.5047 & 0.7433 & 1.0635$^{***}$ & 0.5580$^{***}$ & 5.8444 & 6.2040 & 2.1837$^{**}$ & 2.2506 & $-$0.4209 \\ 
  & (10.5707) & (6.1541) & (2.7046) & (0.3030) & (0.1999) & (3.5885) & (7.1416) & (1.0968) & (2.3797) & (0.3256) \\ 
  & & & & & & & & & & \\ 
 Average Wage & 0.2428$^{**}$ & 0.2220$^{***}$ & 0.0850$^{***}$ & $-$0.0018 & 0.0041 & $-$0.0203 & 0.1439$^{*}$ & 0.0038 & 0.0661$^{**}$ & 0.0050 \\ 
  & (0.1162) & (0.0618) & (0.0248) & (0.0021) & (0.0029) & (0.0263) & (0.0790) & (0.0086) & (0.0263) & (0.0037) \\ 
  & & & & & & & & & & \\ 
 Oil Price & $-$0.0345$^{**}$ & $-$0.0805$^{***}$ & $-$0.0012 & 0.0005 & $-$0.0001 & 0.0289$^{**}$ & $-$0.0433$^{***}$ & 0.0072 & 0.0008 & $-$0.0014 \\ 
  & (0.0143) & (0.0210) & (0.0060) & (0.0012) & (0.0009) & (0.0143) & (0.0096) & (0.0050) & (0.0075) & (0.0011) \\ 
  & & & & & & & & & & \\ 
 GDP per Capita & $-$0.0008 & $-$0.0052 & 0.0007 & 0.0002 & 0.00004 & 0.0035$^{**}$ & $-$0.0004 & 0.0003 & 0.0002 & 0.0002 \\ 
  & (0.0026) & (0.0040) & (0.0008) & (0.0002) & (0.0001) & (0.0017) & (0.0025) & (0.0003) & (0.0006) & (0.0002) \\ 
  & & & & & & & & & & \\ 
 Inflation Rate & 2.4717$^{*}$ & 2.0453 & 0.5946 & $-$0.2218 & 0.0919 & 0.4443 & 2.2318$^{*}$ & $-$0.1226 & $-$0.0550 & $-$0.2227$^{*}$ \\ 
  & (1.3773) & (1.5799) & (0.6899) & (0.1667) & (0.1195) & (1.8419) & (1.3167) & (0.4182) & (0.6204) & (0.1251) \\ 
  & & & & & & & & & & \\ 
 Corruption Index & $-$41.7691 & $-$55.3200 & 7.7657 & 3.5124 & 0.7639 & 18.0214 & $-$45.7457 & 4.9879 & 4.3624 & $-$0.5887 \\ 
  & (31.1907) & (57.9895) & (17.0326) & (2.7032) & (1.2438) & (27.0645) & (31.7703) & (8.5768) & (8.5397) & (2.8701) \\ 
  & & & & & & & & & & \\ 
 Stability Index & 29.9609 & 22.2890 & 6.5295 & $-$0.8532 & 0.9412 & 4.8123 & 28.8083$^{*}$ & 1.0526 & $-$1.1133 & $-$0.6255 \\ 
  & (21.8045) & (19.0913) & (12.4585) & (2.3066) & (0.8356) & (8.8980) & (16.3032) & (5.3891) & (6.1377) & (0.8355) \\ 
  & & & & & & & & & & \\ 
 Rule of Law Index & $-$31.9327 & $-$17.1308 & $-$3.1501 & $-$0.6833 & $-$2.5373 & $-$24.2609 & $-$38.3563 & $-$6.0859 & $-$1.0141 & $-$2.8653 \\ 
  & (28.6649) & (34.3069) & (13.8111) & (5.6023) & (2.8207) & (27.2863) & (27.4561) & (5.9984) & (8.1939) & (1.8894) \\ 
  & & & & & & & & & & \\ 
 Trade Exports & $-$3.8204$^{*}$ & 0.9095 & $-$2.6045$^{***}$ & $-$0.1439 & $-$0.2751$^{**}$ & $-$1.3547 & $-$3.0446$^{*}$ & $-$0.7062 & $-$0.7092 & $-$0.3871 \\ 
  & (2.1536) & (2.6750) & (0.8743) & (0.2310) & (0.1241) & (1.5563) & (1.6371) & (0.4314) & (0.7394) & (0.3075) \\ 
  & & & & & & & & & & \\ 
 Trade Imports & 4.8937$^{**}$ & 4.4906$^{*}$ & 2.2017$^{**}$ & 0.0011 & 0.0232 & $-$1.8789 & 4.6835$^{***}$ & 0.3215 & 0.9043 & 0.3731 \\ 
  & (2.0143) & (2.3825) & (0.8539) & (0.2712) & (0.1237) & (1.8877) & (1.4634) & (0.3807) & (0.7902) & (0.2829) \\ 
  & & & & & & & & & & \\ 
 Outward FDI & 0.7128$^{*}$ & 2.0550$^{***}$ & $-$0.0109 & $-$0.0674$^{**}$ & $-$0.0083 & $-$2.8904$^{***}$ & 0.2699 & 0.2632$^{**}$ & $-$0.0233 & 0.2613 \\ 
  & (0.3664) & (0.7665) & (0.2474) & (0.0293) & (0.0168) & (0.7054) & (0.3322) & (0.1100) & (0.0512) & (0.1703) \\ 
  & & & & & & & & & & \\ 
 Inward FDI & 0.4024 & 0.2644 & 0.2110$^{**}$ & 0.0059 & 0.0330$^{*}$ & 0.8692$^{***}$ & 0.2220 & 0.0891 & $-$0.1040$^{**}$ & $-$0.0233 \\ 
  & (0.3387) & (0.4543) & (0.0987) & (0.0321) & (0.0173) & (0.2771) & (0.3102) & (0.1066) & (0.0481) & (0.0199) \\ 
  & & & & & & & & & & \\ 
\hline \\[-1.8ex] 
Observations & 274 & 241 & 240 & 206 & 241 & 237 & 275 & 278 & 277 & 205 \\ 
R-Squared & 0.1952 & 0.2865 & 0.2064 & 0.0527 & 0.1402 & 0.3759 & 0.1347 & 0.0796 & 0.1480 & 0.2024 \\ 
Serial Correlation Test & 2.5409 & 11.9396$^{***}$ & 0.2782 & 0.009618 & 5.0763$^{**}$ & 0.9176 & 3.6495$^{*}$ & 0.3711 & 0.7340 & 20.2576$^{***}$ \\ 
Heteroscedasticity Test & 57.5123$^{***}$ & 20.2476$^{**}$ & 9.8768 & 39.5827$^{***}$ & 12.0507 & 52.8932$^{***}$ & 40.9034$^{***}$ & 26.2554$^{***}$ & 9.4661 & 13.6443 \\ 
\hline 
\hline \\[-1.8ex] 
\multicolumn{11}{r}{$^{*}$p$<$0.1; $^{**}$p$<$0.05; $^{***}$p$<$0.01} \\ 
\end{tabular} 
\end{sidewaystable}

\section{Conclusion}
\label{sec:con}

Research on both the type and size of government involvement in the innovation process seems to be in early stages despite significant body of already existing literature. Both theory and practical experience gives contradicting information about many important issues. In our paper, we have suggested that private investment in R\&D will happen only if it is efficient, i.e.\ there is income to balance the cost of it. The government investment on the other hand will continue without this condition.

Our focus was dependence of the investment to R\&D on the economic environment in a situation of major governmental involvement in the new millennia. We tried to answer the question whether the market conditions of relative prices of factors of production still hold sway as motivating element for the spur to innovation as suggested by our alternative Hicks hypothesis. Or, if original Hicks hypothesis apply, or any other pattern can be found.

We work with the alternative Hicks hypothesis formulated as follows: innovation is spurred by an increase in the relative price of one factor of production compensated by a decrease in relative price of another factor of production. This hypothesis was proven for part of the investment to R\&D which was either funded or performed by business enterprises. In this case, we found that market sources play a major role. The finding is statistically significant and robust. Under market conditions innovation is only possible when business not only has the need for innovation (increased relative price of factor of production) but also has the resources to conduct it (decrease in relative price of another factor of production) or innovation can bring about such a decrease in form of for example labour and resources-saving techniques with minimal lag.

Investment to R\&D funded by government and investment performed by universities fell under original Hicks hypothesis as there seems to be no need for conditions other than increase of relative price of factor of production. Presumably government funding is the force which allows for research to continue without the need for it to provide other continuous source of income. Continuous crowding out of private investment by government funds with its less desirable properties, namely weaker performance, might cause original Hicks hypothesis to be more relevant then before.

\section*{Acknowledgements}
\label{sec:acknow}

We would like to thank Jan Zouhar for his useful comments. We would also like to thank organizers and participants of the 29th Eurasia Business and Economics Society Conference (Lisbon, October 10--12, 2019) for fruitful discussions.

\section*{Funding}
\label{sec:fund}

The work of Vladimír Holý was supported by the Internal Grant Agency of the Prague University of Economics and Business under project F4/27/2020.

%\bibliography{library.bib}
%\bibliographystyle{mynatstyle}

\end{document}